\documentclass[floatfix,fleqn]{iopart}% Physical Review B
\usepackage[sort&compress,square,comma,numbers]{natbib}
\usepackage{amsfonts}
\usepackage{amssymb}
\usepackage{array}   % Align table columns on decimal point
\usepackage{dcolumn} % Align table columns on decimal point
\usepackage{bm}      % bold greek math symbols
\usepackage{hyperref}
\usepackage{longtable}
\usepackage{graphicx} % Include figure files
\usepackage{xcolor}   % \textcolor{red}{...}

\begin{document}
\newcommand{\text}[1]{{\rm {#1}}}
\title{Thermopower for a molecule with vibrational degrees of freedom}

\author{Sergei Kruchinin$^1$ and Thomas Pruschke$^2$}%
\ead{skruchin@i.com.ua }%
\address{$^1$ Bogolyubov Institute for Theoretical Physics,\\
NASU,\ Kiev   03143, Ukraine}
\address{$^2$ 
Institut f\"ur Theoretische Physik, Georg-August-Universit\"at
G\"ottingen, Friedrich-Hund-Platz 1, 37077 G\"ottingen, Germany}

\date{\today}

\begin{abstract}
We propose a simple model to study resonant tunneling through an organic molecule between to conducting leads, taking into account the vibrational modes of the molecule. We solve the model approximately analytically in the weak coupling limit and give explicit expressions for the thermopower and Seebeck coefficient. The behavior of these two quantities is studied as function of model parameters and temperature. For a certain regime of parameters a rather peculiar variation of the thermopower and Seebeck coefficient is observed. 

Although the model is very simple, we expect it to give some nontrivial insight into thermal transport properties through nan-devices. Furthermore, because we can provide an analytical solution, it may eventually serve as benchmark for more advanced  analytical or computational methods.
\end{abstract}
\noindent{\it Keywords\/}: {Nano devices, thermopower, bose-fermi model}
% Use showkeys class option if keyword display desired
%\maketitle
%
\section{Introduction}
In a recent experiment, Reddy et al.\ \cite{Reddy_Science_2007} studied transport through an organic molecule (benzenedithiol (BDT)) attached to a scanning-tunneling-microscop (STM) tip and a substrate, generated by a break junction using the STM tip. The authors were able to measure different transport properties, including the thermopower respectively Seebeck coefficient of this setup, which is
\begin{figure}[htb]
\centering
 \includegraphics[width=0.6\textwidth]{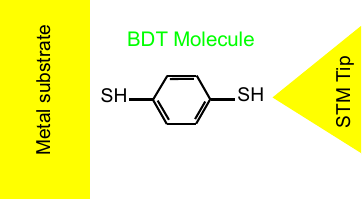}
 \caption{Schematic experimental setup according to \cite{Reddy_Science_2007}. The STM tip is connected to a reservior which maintains a constant tip temperature. The substrate, on the other hand, can be heated, thus creating a temperature gradient across the molecule.}\label{fig:1}
 \end{figure}
is shown schematically in Fig.\ \ref{fig:1}. Quite generally, the idea to contact organic molecules to leads in a controlled fashion is rather fashionable and - if successful - can lead to a variety of interesting applications, ranging from nanoscale computing to quantum computing \cite{applications1,applications2,applications3}. First progress of reproducibly and controlled contacting organic molecules has been made using carbon nanotubes \cite{nanotubes,suspended1,suspended2,suspended3}, but similar attempts with smaller molecules usually suffer from several problems, among them the actual question of how the contact is established and how one can control its quality \cite{problems1,problems2,problems3}.

It is well known, that  the physics of nanoobjects coupled to conducting leads is highly nontrivial (for an early review see for example \cite{overview}, more recent aspects are discussed in \cite{overview2}), and already concentrating on electronic degrees of freedom only presents an extreme challenge to both experiment and especially theory. The latter is due to the fact that correlations on these nanostructures can usually not be ignored and hence one is plagued with complicated quantum impurity problems \cite{goldhaber-gordon,cronenwett}, which can be solved only numerically and only in equilibrium and for rather restricted model setups. The situation becomes even worse if one wants to take into account vibrational modes present in complex molecules \cite{von-oppen-0,von-oppen-1,von-oppen-2,ramsak}. Here, even advanced numerical tools rather quickly meet their limits \cite{ramsak}. 

And last, but by no mean least, the conventional experimental probes used in nano-systems are transport properties, in particular by applying a bias voltage or, as in the above-mentioned experiment, a temperature difference across the nano-object. In this situation, the assumption of thermal equilibrium is quickly questionable, and reliable tools to study correlated systems out of thermal equilibrium -- even only the stationary state -- are not yet existing.

Therefore, setting up simple models which preferably are solvable analytically and which allow to study the physical properties also off equilibrium in the stationary state, are still a highly relevant approach to obtain an at least qualitative feeling how various degrees of freedom influence the transport properties of nano-objects. Particularly thermal transport and here thermopower or Seebeck coefficients are an interesting area because possible applications in the area of energy conversion or storage have gained strong renewed interest over the past decade. However, in this area relatively little theoretical work on thermopower for mesoscopic systems exists to our knowledge so far \cite{ermakov}.

The situation in an experimental setup like the one shown in Fig.\ \ref{fig:1} has the advantage, that one can, for example by stretching the bond through removing the tip, control the coupling between the leads and the molecule very sensitively and actually shift the system into the weak coupling regime. Here, theoretical calculations become much easier. For our purpose, we will therefore assume that the system is in this regime, and suggest a very simple model which will allow us to see how the interplay between electronic and vibrational degrees of freedom influence transport properties. 

The model setup and general expressions for thermopower and possible situations will be presented in the next section. The central result, viz analytical expressions for the thermopower and Seebeck coefficient will be derivedin section \ref{sec:results}, and a summary will conclude the paper.
\section{Setup and model}
We study an organic molecule attached to two metallic leads. The leads are assumed to be in thermal equilibrium but can have different temperatures and chemical potentials. We will take the temperature and chemical potential of the right leads as reference and write $T_R=T$, $T_L=T+\Delta T$ respectively $\mu_R=\mu$, $\mu_L=\mu+\Delta\mu$. In the following, we use $\mu=0$ as reference of energy.

Using first-principle approaches, such a setup has been studied before in a quite general way \cite{Chen_Nano_2003}, concentrating however on the local properties of the molecule. Concentrating on a subset of degrees of freedom allows the use of more advanced analytical and numerical tools and has been applied to several model situations related to experiments mentioned in the introduction \cite{von-oppen-2,ramsak} (see also further references in \cite{ramsak}).

Here, we intend to investigate transport properties, and how local vibration modes influence them. Calculating transport for a general situation becomes rather quickly a cumbersome task. Therefore, in order to be able to gain some insight into the physics of the problem, while at the same time having as little parameters as possible, we concentrate her on a very simple model. Furthermore, restricting the study to a bare-bone model also allows for an explicit analytical solution. Such a solution is rather valuable in several respect, for example as possible benchmark for more advanced methods.

To this end we  assume that the HOMO of the molecule has a certain energy $\epsilon_0$ below the chemical potential of the leads, but the LUMO 
\begin{figure}[htb]
\centering
 \includegraphics[width=0.7\textwidth]{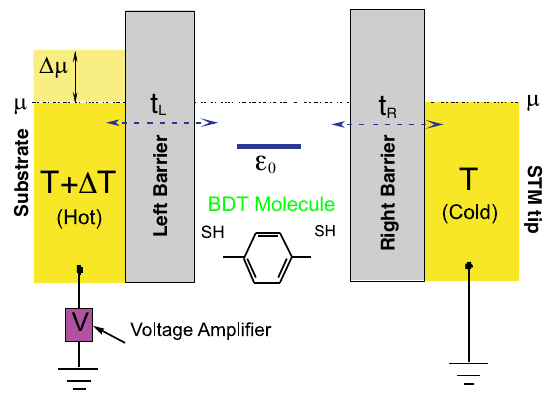}
 \caption{Schematic setup of the model. We consider only one relevant orbital of the BDT molecule, which we assume to be the HOMO. The coupling between the substrate respectively the STM tip are modeled as tunneling barriers.}\label{fig:2}
 \end{figure}
well above it, so that we can neglect it. Also, all other electronic states of the molecule are assumed to be well separated, too. The model motivated above is shown schematically in Fig.~\ref{fig:2}, where we also explicitly included the various model parameters.

Up to now the model is the standard setup to describe resonant transport through a mesoscopic system like molecules or quantum dots. Here, however, we also allow for vibrational modes of the molecule, which we describe in a harmonic approximation, and which couple to the electronic degrees of freedom via a Holstein term. Although the calculation can be performed for several phonon modes, we concentrate here on the case of one single mode with frequency $\omega_0$. This is for example permissible when one studies  low temperature, where only the lowest-lying modes play a role anyway. Finally, the molecule is coupled to the leads via tunneling barriers.

Thus, the Hamiltonian for our model to describe the experimental setup from Fig.~\ref{fig:1} reads
 \begin{eqnarray}
 \label{eq:Htot}
 H &=& H_\text{L}+H_\text{M}+H_\text{L-M}\\
 \label{eq:HLeads}
 H_\text{L} &=& \sum\limits_{{\bf k},\alpha=L,R}\varepsilon_{\alpha,{\bf k}}c_{\alpha,\bf k}^\dagger c_{\alpha,{\bf k}}^{\phantom{\dagger}}\\
 \label{eq:HMol}
H_\text{M} &=&  \epsilon_0\,d^\dagger d+\omega_0\,b^\dagger b+\lambda\omega_0\,d^\dagger d\left(b^\dagger+b\right)\\
H_\text{L-M} &=&
 \sum\limits_{\alpha=L,R}\left(t_\alpha c^\dagger_{\alpha,B}d+\text{h.c.}\right)
 \label{eq:HLM}
 \end{eqnarray}
 where we use standard second quantized notation,  $\lambda$ denotes the dimensionless coupling between the electronic and phononic modes, and $t_\alpha$ the tunneling matrix elements to the leads. The operator $c_{\alpha,B}^{(\dagger)}$ denotes the projection of the band states to the barrier. 
 
Note that we did not include the spin degrees of freedom of the electrons in the model (\ref{eq:Htot}). Although this is at first glance a rather crude approximation, we want to employ it nevertheless for two reasons. First, including the spin makes already the molecular problem highly nontrivial and would prevent the semi-analytical solution we will discuss in the section about the results. Second, we are presently aiming at a more qualitative insight into the influence of vibrational modes on transport effects. To this end we believe that the additional complications due to additional electronic degrees of freedom will surely alter the results quantitatively, but possibly not strongly om a qualitative level. Nevertheless, calculations based on more realistic models are in progress.

For our simple setup we can use the Meir-Wingreen formula \cite{Meir_Wingreen} to calculate the electric current as
 \begin{equation}\label{eq:MeirWingreen}
 J = J_0\int d\omega\left[f_L(\omega)-f_R(\omega)\right] \rho_M(\omega,\Delta\mu, \Delta T)
 \end{equation}
 where $\rho_M(\omega,\Delta\mu,\Delta T)$ denotes the electronic density of states on the molecule in the presence of a potential difference $\Delta V$ and a temperature difference $\Delta T$ between the leads, and 
 $$
 f_\alpha(\omega)=\frac{1}{1+\exp\left(\frac{\omega-\mu_\alpha}{k_\text{B}T_\alpha}\right)}
 $$
 the Fermi functions of the left and right lead, respectively. The quantity $J_0$ finally collects the coupling parameters and the natural constants like charge and Planck's constant.
 
\begin{figure}[htb]
\begin{center}
 \includegraphics[width=0.49\textwidth]{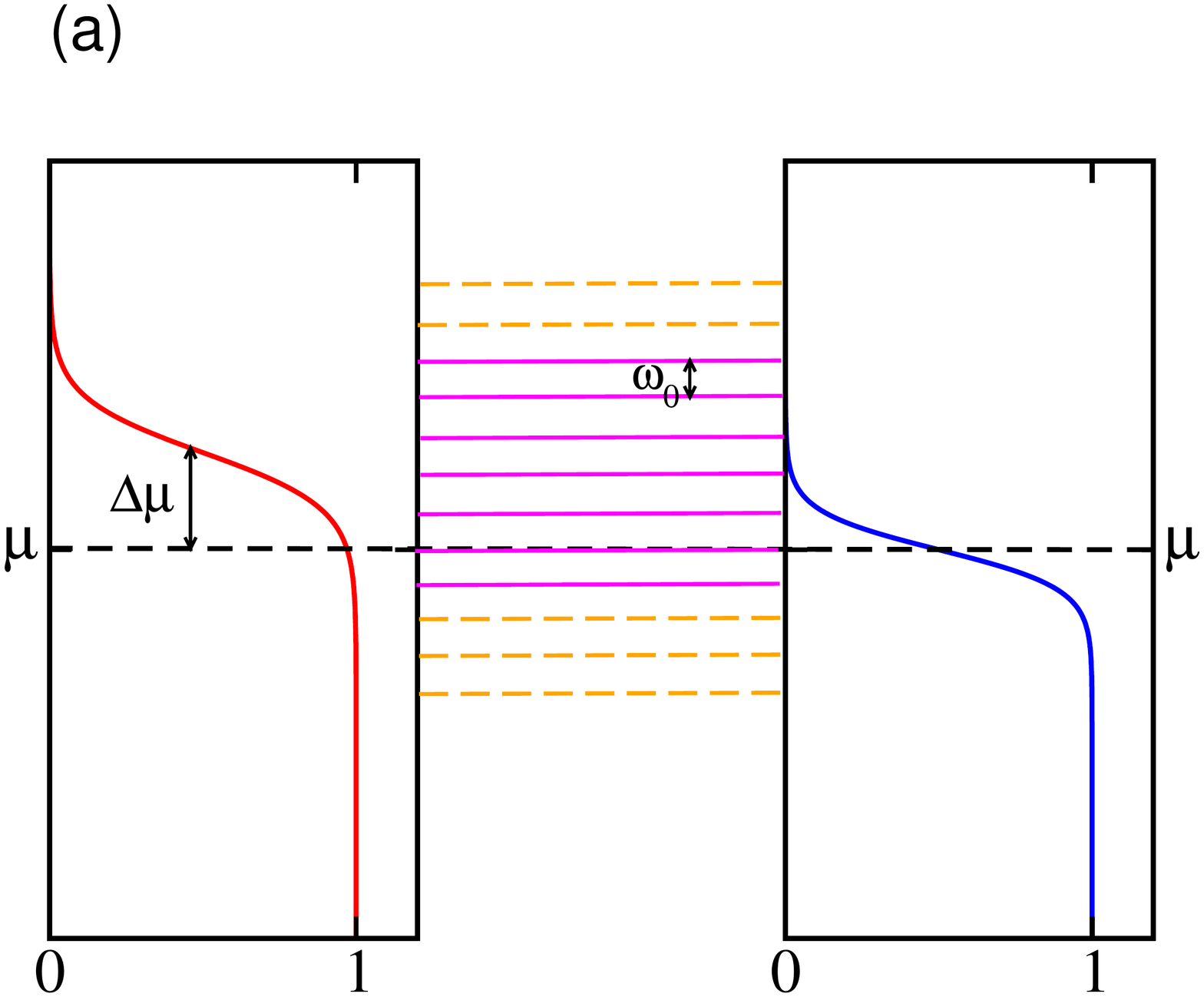}
 \hfill
 \includegraphics[width=0.49\textwidth]{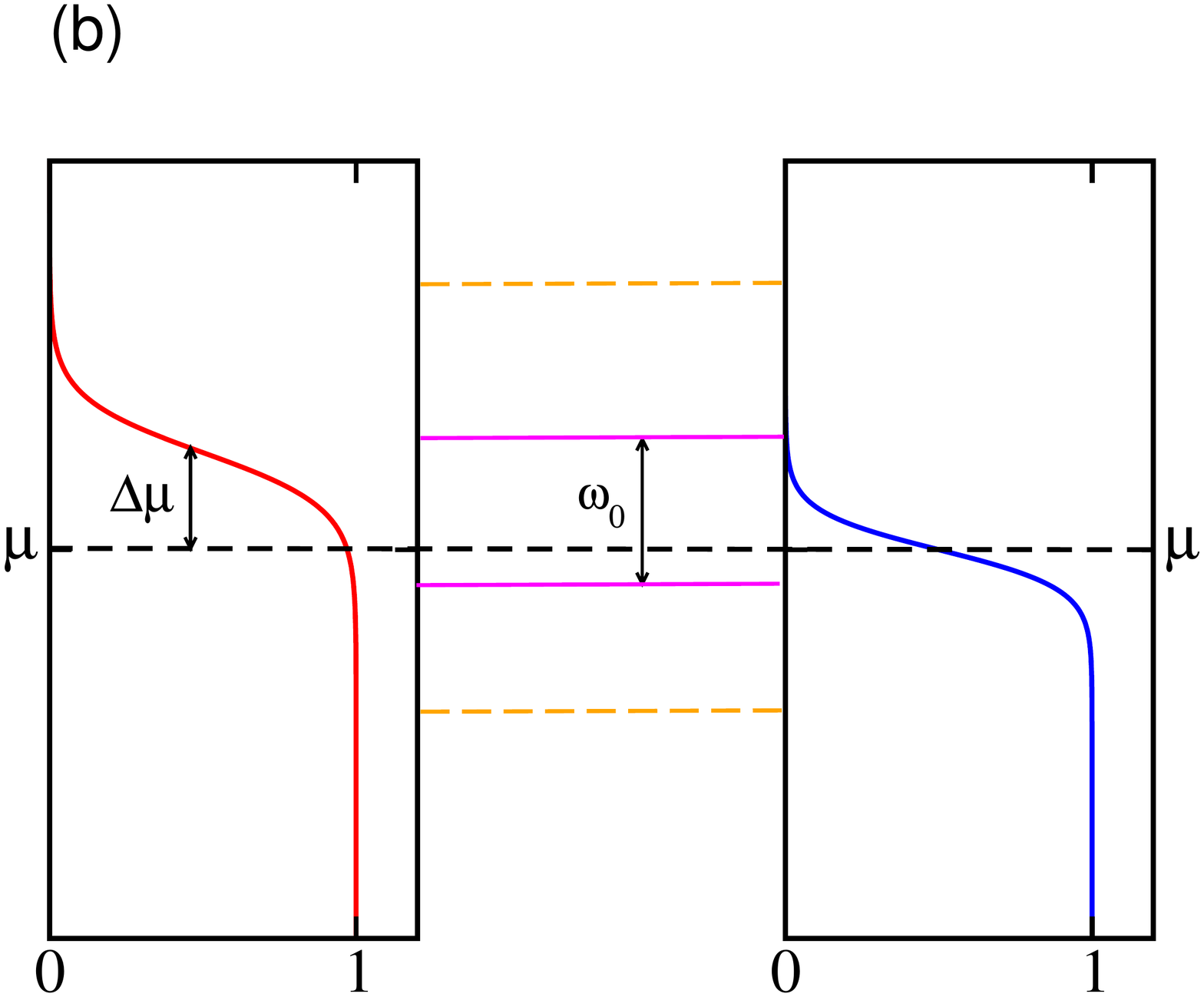}
 \end{center}
 \caption{Schematic representation of the different situations arising from Eq.~(\ref{eq:condition_full}). The left and right boxes denote the left and right leads, with the Fermi functions as full lines. In the left lead, we included some $\Delta\mu$ and $\Delta T$. In (a) there are several polaronic modes which lie in the Fermi window, while in (b) the temperature is much smaller than the level splitting, resulting in effectively only two modes within the Fermi window.}\label{fig:3}
 \end{figure}

 Calculating the DOS in the presence of the leads for arbitrary coupling $t_\alpha$, $\Delta\mu$ and $\Delta T$ is not possible. However, in the case $t_\alpha\to0$ (weak coupling), one can approximate $\rho_M(\omega,\Delta\mu, \Delta T)$ by the result for the isolated molecule, which reads \cite{mahan}
 \begin{equation}\label{eq:DOS_wk}
 \rho_M(\omega,\Delta\mu,\Delta T)=\frac{2\pi}{e^\lambda}\sum\limits_{l=-\infty}^\infty\frac{\lambda^{|l|}}{|l|!}\,\delta(\omega-\epsilon_0-(l+\lambda)\omega_0)
 \end{equation}
 and does not depend on $\Delta\mu$ and $\Delta T$ explicitely.
 When we insert expression (\ref{eq:DOS_wk}) into the Meir-Wingreen formula (\ref{eq:MeirWingreen}), the expression for the current becomes
% \begin{widetext}
 %\begin{equation}
 $$
 J=\frac{2\pi J_0}{e^\lambda}\sum\limits_{l=-\infty}^\infty\frac{g^{|l|}}{|l|!}\,
 %\bigl[\!\begin{array}[t]{l}
 %\displaystyle 
 \left[f_L(\epsilon_0+(l+\lambda)\omega_0)-\\[2mm]
 %\displaystyle 
 f_R(\epsilon_0+(l+\lambda)\omega_0)\right]
 %\bigr]\end{array}
 $$
 %\end{equation}
 %\end{widetext}
Since we are interested in the thermopower, the current has to be zero, i.e.\ for a given temperature difference $\Delta T$ between left and right lead we have to adjust $\Delta\mu$ such that
 \begin{eqnarray}
 \sum\limits_{l=-\infty}^\infty\frac{\lambda^{|l|}}{|l|!}
% &\bigl[&%
\left[ f_L(\epsilon_0+(l+\lambda)\omega_0)-
%\nonumber\\
%&&
\label{eq:condition_full}
%\hspace{-2mm} 
f_R(\epsilon_0+(l+\lambda)\omega_0)\right]=0
 \end{eqnarray}
In general, we are interested in low temperatures, i.e.\ typically we will have $|\epsilon_0|$, $\omega_0\gg k_\text{B}T$. In this case, only a certain range of values of $l\in[l_\text{min},l_\text{max}]$, which have the property that $\epsilon_0+(l+\lambda)\omega_0={\cal O}(k_\text{B}T)$, will actually be important, because for the other terms the Fermi functions are either both zero or both one, i.e.\ cancel in the difference. This situation is schematically depicted in Fig.~\ref{fig:3}a.

Let us take this idea to the extreme and assume, that $T$ is low enough such that only one vibrational mode with index $l_0$ will contribute, see the schematic representation in Fig.~\ref{fig:3}b. The necessary conditions for transport in this case are
\begin{eqnarray*}
\epsilon_0+(l_0+\lambda)\omega_0 &<& 0\\
0\le \tilde{\epsilon}_0:=\epsilon_0+(l_0+1+\lambda)\omega_0&<&\omega_0
\end{eqnarray*}
from which $l_0$ can be determined. The equation (\ref{eq:condition_full}) then simplifies to
%\begin{widetext}
\begin{equation}\label{eq:condition_simple}
\begin{array}[b]{l}\displaystyle
\frac{1}{1+\exp\left(\frac{\tilde\epsilon_0-\omega_0-\Delta\mu}{k_\text{B}(T+\Delta T)}\right)}
-
\frac{1}{1+\exp\left(\frac{\tilde\epsilon_0-\omega_0}{k_\text{B}T}\right)}
+\\[5mm]
\displaystyle\frac{\lambda}{l_0+1}\left[
\frac{1}{1+\exp\left(\frac{\tilde\epsilon_0-\Delta\mu}{k_\text{B}(T+\Delta T)}\right)}
-
\frac{1}{1+\exp\left(\frac{\tilde\epsilon_0}{k_\text{B}T}\right)}
\right]=0\end{array}
\end{equation}
%\end{widetext}
This equation can be easily solved for $\Delta\mu$ for given model parameters and $\Delta T$. Although this is strictly speaking not  the case for the experimental setup in Fig.~\ref{fig:1}, we want to note that in nano devices $\epsilon_0$  can often be controlled through a gate voltage, $\epsilon_0=-eV_G$, and consequently the active polaronic mode respectively $l_0$ is controllable experimentally. 
\section{Results\label{sec:results}}
Without vibrational modes, the thermoelectric response of the model (\ref{eq:Htot}) has been studied before by Ermakov~et al.\ \cite{ermakov} Within their model, the authors observed a linear relation between the temperature difference and thermovoltage. In the following we want to try to understand what modifications arise through the presence of polaronic modes in the molecule.

In the present paper we want to concentrate on the situation shown in Fig.~\ref{fig:3}b. Here we can 
use the fact that  $\omega_0\gg k_\text{B}T$ and try to solve equation (\ref{eq:condition_simple}) approximately analytically and provide explicit expressions for $\Delta\mu$ and the Seebeck coefficient as function of model parameters and temperature.

The limiting cases $\tilde\epsilon_0\to0$ and $\tilde\epsilon_0\to\omega_0$ turn out to be uninteresting, because in the first case the combination$f_L(\tilde\epsilon_0-\omega_0)-f_R(\tilde\epsilon_0-\omega_0)$ vanishes, while in the second $f_L(\tilde\epsilon_0)-f_R(\tilde\epsilon_0)\to0$. In both cases one thus recovers the solution by Ermakov et al.\cite{ermakov} in the limit when the level position is close to the chemical potential.

The more interesting case therefore is $\tilde\epsilon_0$  finite and reasonably far away from the chemical potential. Then we can furthermore assume $\tilde\epsilon_0\gg k_\text{B}T$ as well as $|\tilde\epsilon_0-\omega_0|\gg k_\text{B}T$, and replace the Fermi functions by $f(x)\approx e^{-\beta x}$, if $x>0$, respectively $f(x)\approx1-e^{\beta x}$, if $x<0$. Inserting these approximations into Eq.\ (\ref{eq:condition_simple}) one obtains a quadratic equation for $e^{\frac{\Delta\mu}{k_\text{B}(T+\Delta T)}}$. From the two solutions one must pick the one that has the correct limit 
$$
\Delta\mu=\left(\omega_0-\epsilon_0\right)\frac{\Delta T}{T}
$$
as $\lambda\to0$. The solution fulfilling this requirement reads
%\begin{widetext}
\begin{eqnarray}
\hspace{-10mm}
\Delta\mu&\approx&
(T+\Delta T) \ln\frac{e^{-\frac{\tilde\epsilon_0}{k_\text{B}T}}}{\frac{2
   \lambda}{l_0+1}}\Biggl[
   e^{\frac{\tilde\epsilon_0}{k_\text{B}(T+\Delta T)}}
   \left(\frac{\lambda}{l_0+1} -e^{\frac{2\tilde\epsilon_0-\omega_0}{k_\text{B}T}}\right)+\nonumber\\[7mm]
   \label{eq:sol_approx}\hspace{-10mm}
   &&\sqrt{e^{\frac{2
   \tilde\epsilon_0}{k_\text{B}(T+\Delta T)}} \left(e^{\frac{2\tilde\epsilon_0-\omega_0}{k_\text{B}T}}-\frac{\lambda}{l_0+1} \right)^2+\frac{4
   \lambda}{l_0+1}  e^{\frac{2\tilde\epsilon_0-\omega_0}{k_\text{B}(T+\Delta T)}+\frac{2 \tilde\epsilon_0}{k_\text{B}T}}}\;\Biggr]
\end{eqnarray}
%\end{widetext}
%   
%\begin{eqnarray*}
%\Delta\mu &=& 
%\left(\omega_0-\epsilon_0\right)\frac{\Delta T}{T}+\\
%&&k_\text{B}\left(T+\Delta T\right)\ln\frac{1+\frac{\lambda}{l_0+1}e^{-\frac{\omega_0}{k_\text{B}T}}}{1+\frac%{\lambda}{l_0+1}e^{-\frac{\omega_0}{k_\text{B}(T+\Delta T)}}}
%\left(T+\Delta T\right)\ln\frac{e^{-\frac{\epsilon_0\Delta T}{T^2+T\Delta T}}\left(e^{\frac{\omega_0}{T}}+\frac{\lambda}{l_0+1}\right)}{e^{\frac{\omega_0}{T+\Delta T}}+\frac{\lambda}{l_0+1}}
%\end{eqnarray*}
Since this expression is rather hard to interpret, we inspect it behavior for small $\lambda$. Taylor expanding it to first order results in
$$
\Delta\mu\approx
(\omega_0-\tilde\epsilon_0)\frac{\Delta T}{T}-\frac{\lambda(T+\Delta T)}{l_0+1}e^{\frac{\omega_0-2\tilde\epsilon_0}{k_\text{B}T}}\left[e^{\frac{\omega_0\Delta T}{k_\text{B}T(T+\Delta T)}}-1\right]
$$
%This expression reduces to the result by Ermakov et al.\cite{ermakov} in the limit $\lambda\to0$.
%Within the same level of approximation $\omega_0\gg k_\text{B}T$, the logarithm can be expanded and one obtains
%\begin{eqnarray*}
%\Delta\mu&\approx& \left(\omega_0-\epsilon_0\right)\frac{\Delta T}{T}+\\
%&&\frac{k_\text{B}\left(T+\Delta T\right)\lambda}{l_0+1}
%\left[e^{-\frac{\omega_0}{k_\text{B}T}}-e^{-\frac{\omega_0}{k_\text{B}(T+\Delta T)}}\right]\\
%&=&
%\left(\omega_0-\epsilon_0\right)\frac{\Delta T}{T}-\\
%&&\frac{k_\text{B}\left(T+\Delta T\right)\lambda}{l_0+1}e^{-\frac{\omega_0}{k_\text{B}T}}
%\left[e^{\frac{\omega_0\Delta T}{k_\text{B}T(T+\Delta T)}}-1\right]\\
%\end{eqnarray*}
%The first thing to note is that the vibrational part tends to decrease $\Delta\mu$. However, within the assumptions made we only find an exponentially suppressed contribution, which means that the regime of very low temperatures is uninteresting regarding the influence of molecular vibrations.
The first term is identical to the result by Ermakov et al.\ \cite{ermakov}. The correction is negative, i.e.\ the vibrations tend to initially reduce the thermopower. Note that the strenght of this effect depends on the relation between $\tilde\epsilon_0$ and $\omega_0$: If $\omega_0-2\tilde\epsilon_0<0$, it will be exponentially suppressed, while it is actually stronlgy enhanced for $\omega_0-2\tilde\epsilon_0>0$.

\begin{figure}[htb]
\centering
\includegraphics[width=0.8\textwidth]{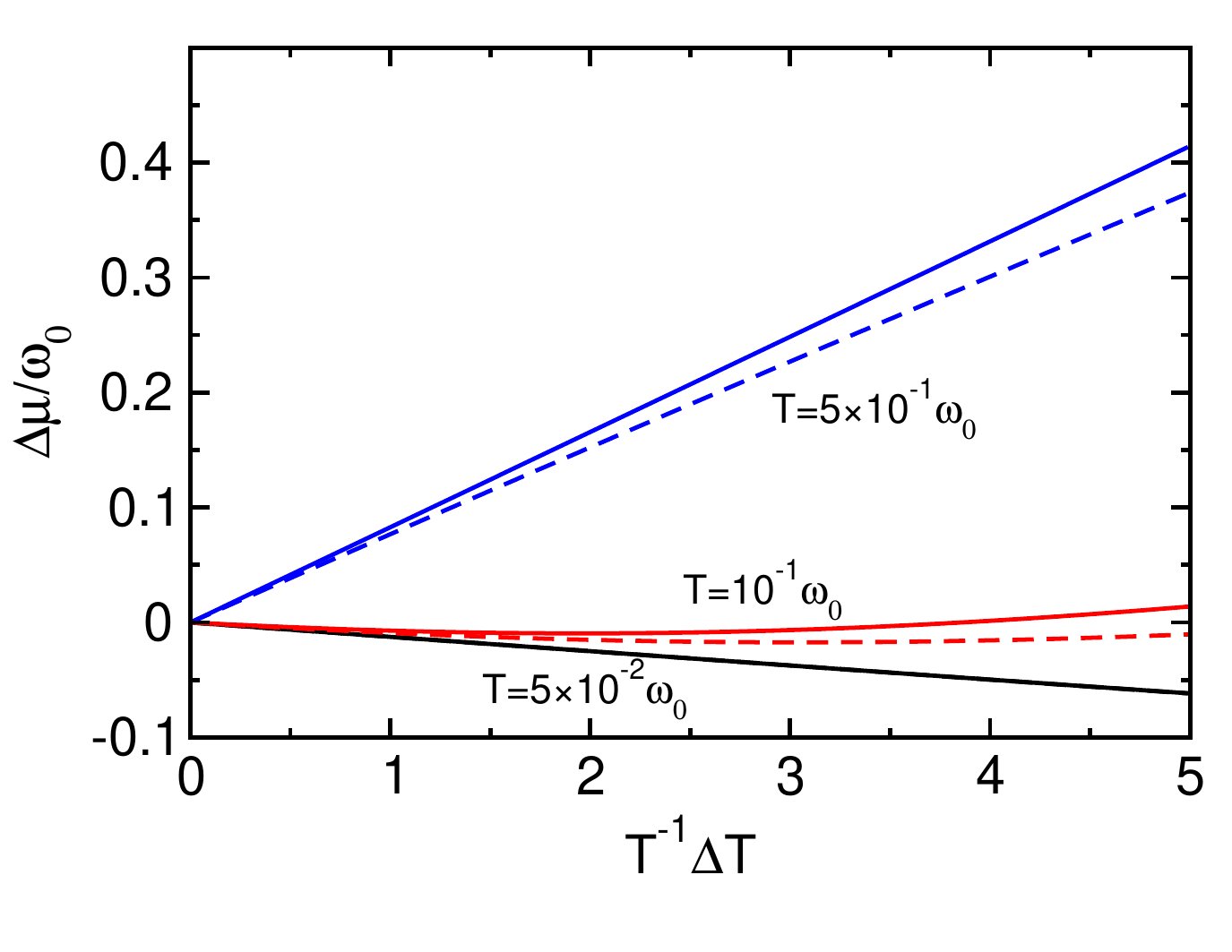}
\caption{$\Delta\mu$ vs.\ $T^{-1}\Delta T$ for various values of $T$. Other parameters see text. The dashed curves are the results from the approximate formula (\ref{eq:sol_approx}).}\label{fig:4}
\end{figure}

Let us see in to what extent the approximate solution (\ref{eq:sol_approx})  is reliable. To this end we 
solve Eq.~(\ref{eq:condition_simple}) numerically for $\Delta\mu$. As generic parameter set we chose $\tilde\epsilon_0=\omega_0/8$ and $\lambda/(l_0+1)=1/40$. Due to the exponential structure a numerical solution fails for too small $T$.  Using multi-precision arithmetic, one can extend the range of solvability slightly. We checked the stability of the root finding algorithm by comparing the results for $\Delta\mu$ at $\lambda\to0$ with the analytical result and found that stable solutions can be obtained for $T\ge 5\cdot10^{-2}\omega_0$ using a precision of 100 digits.   The resulting function $\Delta\mu(\Delta T)$ for $T=5\cdot10^{-2}\omega_0$, $10^{-1}\omega_0$ and $\omega_0/2$ is shown in Fig.\ \ref{fig:4} as full lines,
together with the approximate functions as dashed lines.
For low $T$ we observe a linear dependence with negative coefficient, while at high $T$ the sign of the thermopower reverses. For intermediate $T$, however, one can see a rather peculiar sign change of $\Delta\mu$ as function of $\Delta T$. Changing the model parameter does not influence this qualitative behavior as long as $\omega_0-2\tilde\epsilon_0>0$. For example, decreasing $\tilde\epsilon_0$ moves the zero in $\Delta\mu$ towards smaller values of $\Delta T$, but also decreases the magnitude of the effect. Increasing $\lambda$ produces a similar behavior. However, when $\omega_0-2\tilde\epsilon_0<0$ the features due to the vibrations vanish and one recovers more or less the behavior for $\lambda\to0$, i.e.\ linear with slope $(\tilde\epsilon_0-\omega_0)/T>0$.

The curves resulting from the approximate formula give, as expected from the assumptions behind this analytical approximation, an accurate description for low temperatures, while for higher temperatures deviations become more pronounced. However, the qualitative behavior is well reproduced in all cases and we conclude that the analytical function represents  a reasonable approximation for higher temperatures, too.

\begin{figure}[htb]
\centering
\includegraphics[width=0.8\textwidth,clip]{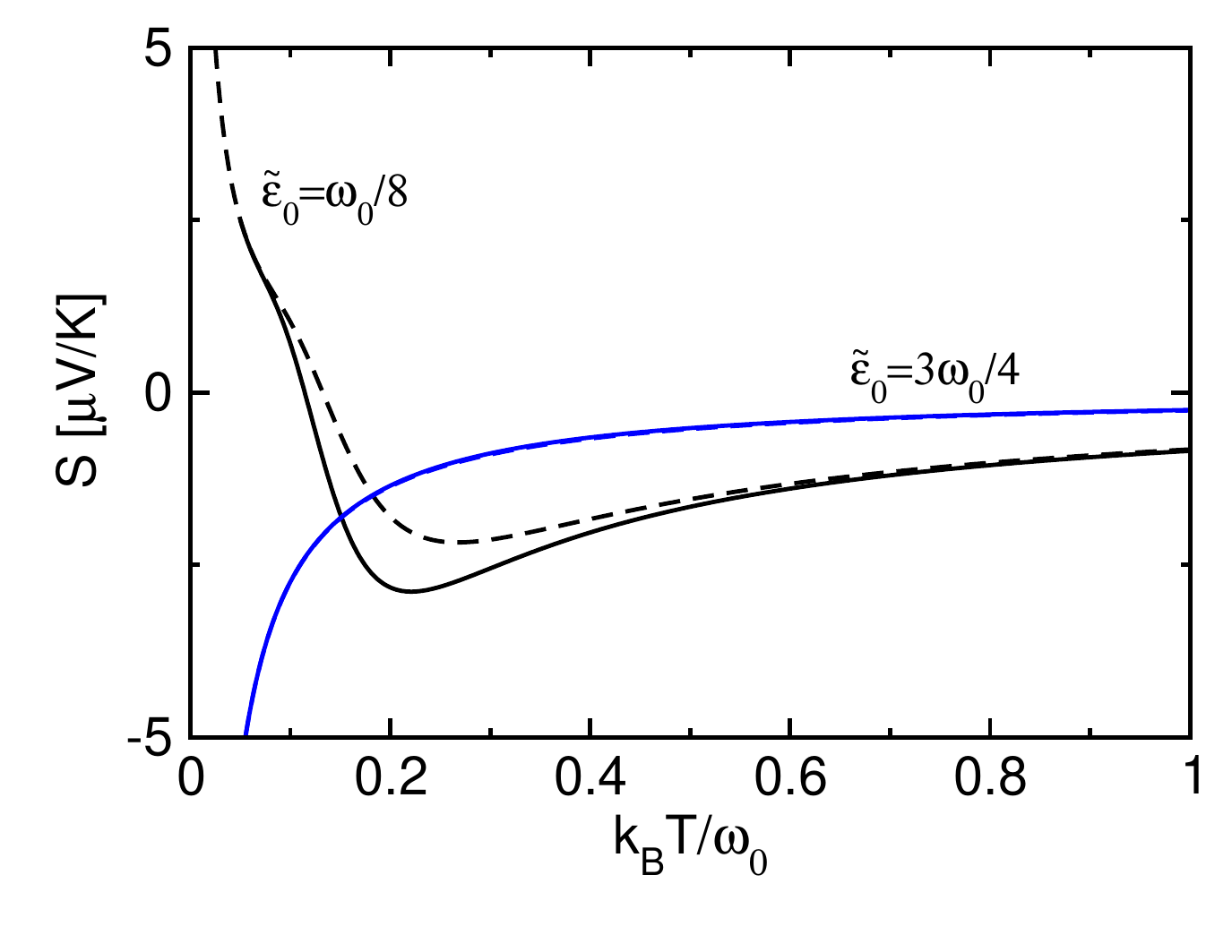}
\caption{Seebeck coefficient $S=-\left.\frac{\Delta V}{\Delta T}\right|_{\Delta T\to0}$ as function of $T$ for two characeristic values of $\tilde\epsilon_0$. Other parameters as in Fig.\ \ref{fig:4}. The dashed curves are the result from the approximate formula (\ref{eq:seebeck_approx}).}\label{fig:5}
\end{figure}

Finally, one can look at the initial slope of $\Delta\mu(\Delta T\to0)$, which, when divided by the electron charge, is the negative Seebeck coefficient $S$ for the junction. The result as function of temperature for the same model parameters used in Fig.\ \ref{fig:4} is shown in Fig.\ \ref{fig:5}.
Again, we include the result ($e_0$ denotes the electric charge)
\begin{equation}\label{eq:seebeck_approx}
S\approx\frac{k_\text{B}}{e_0}\,\frac{1}{k_\text{B}T}\left[\tilde\epsilon_0-
\frac{\omega_0}{1+\frac{\lambda}{l_0+1}  e^{\frac{\omega_0-2 \tilde\epsilon_0}{k_\text{B}T}}}\right]\vspace{8mm}
\end{equation}
derived from the approximate formula (\ref{eq:sol_approx}) as dashed curve in Fig.\ \ref{fig:5}. It apparently 
describes the overall behavior quite well, becoming asymptotically exact for low $T$. Interestingly it also provides the correct asymptotic for high $T$. Note that the unit of the thermopower is, as usual, determined by the universal ratio $k_\text{B}/e_0$ and is hence not model or parameter dependent. As already noted in the case of $\Delta\mu$, we here, too, obtain a sensible dependence of the Seebeck coefficient on $\tilde\epsilon_0$. When $\omega_0-2\tilde\epsilon_0>0$, it shows strong variation with $T$ due to the denominator in Eq.\ (\ref{eq:seebeck_approx}), while for $\omega_0-2\tilde\epsilon_0<0$ the latter influence is exponentially suppressed and hence one obtains a temperature dependence which is very similar to the case without molecular vibrations. Since one can possibly control the value of $\epsilon_0$ and hence $\tilde\epsilon_0$ experimentally, at least for electrically gated molecules, the described effects and their switching could in principle be observable.

\section{Summary}
A simple model for an organic molecule attached to two conducting leads via tunnel barriers was introduced. Main emphasis was laid on the inclusion of vibration modes for this molecule and to study their impact on transport properties, here especially the thermopower. We derived an analytical approximate formula for the thermopower $\Delta\mu(\Delta T)$ and the Seebeck coefficient and showed that for a rather broad range of model parameters the influence of the molecular oscillations can strongly influence these two quantities. In particular, the thermopower can develop a pronounced extremum and sign changes as function of temperature gradient across the molecule, and the Seebeck coefficient showed pronounced nonmomotonic behavior in this regime. When one shifted the local electronic level of the molecule to larger values, one can observe a transition into a regime where the vibrational modes have negligible influence and the system behaves similar to the case without any elastic degrees of freedom. Since in some experimental setups the local parameters are accurately controllable via e.g.\ gate voltage or the like, we are certain that one can test these theoretical predictions to some extent.

Of course, the simplifications introduced to allow for an analytical solution are rather severe: Electron spin as well as additional molecular levels were neglected, and the vibration modes were restricted to a single mode. Furthermore, for the calculation of the transport only two of the polaron levels were taken into account. The latter approximation can more easily be relaxed, even within the proposed model (see Fig.\ \ref{fig:3}a). 
Similarly, including several vibration modes can be done straightforwardly (see for example \cite{mahan}), too, while taking into account more realistic electronic level structures  will actually  require more refined analytical approaces \cite{von-oppen-0} or a numerical solution of the resulting model for the molecule \cite{ramsak}; and possibly also a reinspection of the .validity of the Meir-Wingreen formula for the current.

In any case, adding these additional and physically important features to the model of the molecule makes the problem harder to treat. At least as long as we stay in the weak-coupling limit, we can expect that one can handle them reasonably within existing numerical approaches, or even, up to the solution of the transport equation, to a large extent analytically. The actual problematic part is hidden in the solution of the transport equation, even in weak coupling. Here, the exponential nature of the Fermi functions renders the numeric root searching extremely awkward and  severely limits the accessible parameter and especially temperature regime (see e.g.\ discussion of Fig.\ \ref{fig:4}). Finally, relaxing the condition of weak coupling to the leads leads to a full-scale many-body problem out of equilibrium, which at present cannot be really attacked with reliable analytical or numerical tools.

Thus, to conclude, we believe that even such an extremely simple model will be of some importance to increase our understanding of properties of nano-systems, and can eventually also serve as a benchmark for more advanced methods to be developed in the future.

%\begin{acknowledgements}
\ack
We would like to acknowledge financial support by the Deutsche Forschungsgemeinschaft
through  project PR 298/12 (T.P. and S.K.) and S.K. the SASIIU of the project Ukraine -- Republic of Korea (S.K.). S.K.\ thanks the Institute for Theoretical Physics of the University of G\"ottingen and T.P. the Bogolyubov Institute for Theoretical Physics of the NASU  in Kiev for their support and hospitality during the respective visits.We also thank Dr.\ Ermakov  and  Dr.\ Zolotovsky  for the fruitful discussions.
%\end{acknowledgements}

\end{document}